\documentclass[a4paper,11pt]{article}
\usepackage{graphics,graphicx}
\usepackage{amsmath,amssymb}
\usepackage{cite}

\usepackage{a4wide}
\usepackage{hyperref}
\usepackage{tikz}
\usetikzlibrary{cd}

\newcommand{\sesubset}{\mathrel{
\rotatebox[origin=c]{135}{\large $\subset$}}
}
\newcommand{\swgogosubset}{\mathrel{
\rotatebox[origin=c]{55}{\large $\subset$}}
}

\begin{document}
\thispagestyle{empty}
\begin{flushright}
\end{flushright}
\vspace{8mm}
\begin{center}
{\LARGE\bf Grade restriction and D-brane transport for a non-abelian GLSM of an elliptic curve}
\end{center}
\vspace{10mm}
\begin{center}
  {\large Johanna Knapp\footnote{{\tt johanna.knapp@unimelb.edu.au}}${}^{\dagger}$}
\end{center}
\vspace{3mm}
\begin{center}
  ${}^{\dagger}$ {\em School of Mathematics and Statistics, The University of Melbourne\\
  Parkville, VIC 3010, Australia}
\end{center}
\vspace{15mm}
\begin{abstract}
  \noindent We discuss a simple model for D-brane transport in non-abelian GLSMs. The model is the elliptic curve version of a non-abelian GLSM introduced by Hori and Tong and has gauge group $U(2)$. It has two geometric phases, both of which describe the same elliptic curve, once realised as a codimension five complete intersection in $G(2,5)$ and once as a determinantal variety. The determinantal phase is strongly coupled with unbroken $SU(2)$. There are two singular points in the moduli space where the theory has a Coulomb branch. Using grade restriction rules, we show how to transport B-branes between the two phases along paths avoiding the singular points. With the help of the GLSM hemisphere partition function we compute analytic continuation matrices and monodromy matrices, confirming results obtained by different methods. \\\\
   \em Prepared for the proceedings of the workshop ``Gauged Linear Sigma Models @30'', held at the Simons Center for Geometry and Physics, May 22-26 2023. 
\end{abstract}
\newpage
\setcounter{tocdepth}{1}
\tableofcontents
\setcounter{footnote}{0}

\section{Introduction}
The seminal work by Herbst, Hori and Page\cite{Herbst:2008jq} solved the problem of transporting B-type D-branes between phases of abelian GLSMs. A complete understanding of categorical equivalences implied by non-abelian GLSMs is still a mostly open issue due to complicating factors that are not present in the abelian case.

One challenge is posed by the phases themselves. While geometric phases in abelian GLSMs are complete intersections in toric varieties, whose geometry and associated derived categories of coherent sheaves are well understood, geometric phases in non-abelian GLSMs are more complicated. Phases where the gauge symmetry of the GLSM is completely broken are still complete intersections but in more general ambient spaces. Early examples include complete intersections in Grassmannians\cite{Hori:2006dk} and in free quotients of toric varieties\cite{Hosono:2011np}. More general constructions have been found\cite{Jockers:2012zr,Gerhardus:2015sla,Hori:2016txh,Caldararu:2017usq,Knapp:2019cih,Chen:2020iyo,Knapp:2021vkm,guo2023glsm} but a systematic treatment such as the combinatorial approach to toric geometry is still missing. It also tends to be more difficult to work with the associated derived categories of coherent sheaves. In contrast to abelian GLSMs, non-abelian GLSMs can also have phases where a non-abelian continuous subgroup of the gauge group is unbroken. In physics terms, this means we have to deal with strong coupling phenomena. Mathematically, unbroken non-abelian gauge symmetry appears to imply that the geometries are non-complete intersections, which means that one loses intuition when it comes to constructing branes with certain properties. In principle, these issues can be avoided by going to the weakly coupled dual theory\cite{Hori:2011pd}. However, this is not an option if the goal is to transport branes between phases of the same GLSM. This makes dealing with strongly coupled phases unavoidable.

Another challenging issue is the more complicated structure of the phase boundaries. In abelian one-parameter GLSMs there is only one singular point at the phase boundary where a Coulomb branch emerges while in non-abelian one-parameter models there can be more than one point. This implies that there is more than one inequivalent class of paths along which a D-brane can be transported from one phase to another. Associated to each path there is a grade restriction rule\cite{Herbst:2008jq} which determines a subcategory of GLSM branes that can be safely transported from one phase to another along the given path. The main challenge is to extract the grade restriction rule from the GLSM. A further task is to identify ``empty branes'' associated to each phase that can be used to replace branes that do not satisfy the grade restriction rules by IR equivalent branes that do.

A canonical example to study these problems is a class of Pfaffian/Grassmannian GLSMs with gauge group $G=U(2)$ first discussed by Hori and Tong\cite{Hori:2006dk}. The two Calabi-Yau $N$-fold phases of these one-parameter GLSMs are a complete intersection in a Grassmannian and a Pfaffian variety. For Calabi-Yau threefolds, the GLSM gives a physics realisation of the connection between two non-birational Calabi-Yaus found by R{\o}dland\cite{Rodland:2000ab}. Equivalences of the associated derived categories of coherent sheaves have been discussed in the mathematics literature (eg.\cite{MR3223878,Addington:2014sla,Donovan:2020jfh}). Some results for the GLSM analysis, including proposals for the grade restriction rule are available\cite{MR3673174}. It is expected that one can deduce the grade-restriction rule from the asymptotics of the GLSM hemisphere partition function\cite{Hori:2013ika,grr}. 

The aim of this note is to give a practical guide on how to transport D-branes in non-abelian GLSMs in a simple example. For this purpose we choose the elliptic curve version of the Pfaffian/Grassmannian GLSM. This model has two singular points in the moduli space and thus there are two classes of grade restriction rules related to two families of inequivalent paths. Making use of the proposed grade-restriction rules\cite{MR3673174,Donovan:2020jfh,grr}, a proposal for empty branes in Grassmannian phases\cite{MR3223878} and the hemisphere partition function, we transport a set of branes from the Grassmannian to the Pfaffian phase. By means of the hemisphere partition function we can compute the analytic continuation matrices along specific paths in the moduli space. This is enough information to compute the monodromy matrices around the two singular points at the phase boundary. We confirm a result\cite{Knapp:2021vkm} obtained by numerical analytic continuation of the mirror periods. The matrices have certain modular properties. Our approach is a generalisation of a discussion \cite{Erkinger:2017aaa} applied in the abelian case, building upon the work of Herbst, Hori and Page\cite{Herbst:2008jq}. It has the advantage that monodromies can be computed using homological algabra and evaluating residue integrals rather than by means of numerical methods.

% simplest example, rodland, 3pts, done in math, in progress in physics, simpler examples 

%Everything is already said in\cite{Herbst:2008jq} and in Will Donovan's paper. example-oriented executive summary of HHP
% cite hhp, generalisation to non-abelian, more intricate windows, showcase in simplest possible non-abelian example, some words about the rodland elliptic curve and cite thorsten, modular properties and analytic continuation - give the matrices, reproduce this using the glsm and hemisphere, generalising what id did with david, advantage of the method is that it is purely algebraic in the sense that 
% aim of this work is not on deriving the concepts but to show how they are applied in practice. in the abelian case, this has been shown in section 10 of hhp
\section{GLSM, hemisphere partition function, and D-brane transport}
% general glsm data in std notation, and rodlang-glsm and phases, pf-operators and periods
We characterise a Calabi-Yau GLSM by the data $(G,V,\rho_V,R,W)$, where $G$ is the gauge group, $V$ is a complex vector space -- the space of chiral matter fields $\phi_i$ ($i=1,\ldots,\mathrm{dim}V$) transforming in the representation $\rho_V:G\rightarrow SL(V)$ whose weights $Q_i^a$ ($a=1,\ldots,\mathrm{rk}(G)$) are the gauge charges of the $\phi_i$. There is a $U(1)$ vector R-symmetry acting in the representation $R$ on the matter fields. We denote the corresponding weights by $R_i$. Furthermore we consider a model with a gauge-invariant superpotential $W\in\mathrm{Sym}V^*$ of R-charge $2$. Models with superpotential have phases that are compact Calabi-Yaus. 

Throughout this note we will focus on a GLSM with gauge group $G=U(2)$ and the following matter content:
\begin{equation}
  \begin{array}{c||c|c}
      \phi&p^1,\ldots, p^5&x_1^a,\ldots,x_5^a\\
      \hline
      U(2)&\mathrm{det}^{-1}&\Box\\
      U(1)_R&2-4\kappa&2\kappa
  \end{array}\qquad\qquad 0<\kappa<\frac{1}{2},\quad a=1,2.
\end{equation}
The $p$-fields are singlet fields transforming in the inverse determinantal representation, the $x$-fields are doublets that transform in the fundamental representation. It is often useful to view $x=x_i^a$ as a $2\times 5$-matrix.

The superpotential is
\begin{equation}
  \label{rodw}
  W=\sum_{i,j,k=1}^5\sum_{a,b=1}^2A^{ij}_kp^kx_i^a\varepsilon_{ab}x_j^b=\sum_{i,j=1}^5A^{ij}(p)[x_ix_j],
\end{equation}
where $\varepsilon_{ab}$ is the Levi-Civit\'a symbol, $A^{ij}(p)$ is a sufficiently generic\cite{Hori:2006dk} antisymmetric matrix linear in the $p$-fields and the baryons $[x_ix_j]$ are the Pl\"ucker coordinates on the Grassmannian $G(2,5)\equiv G(S,V)$ with $S\cong\mathbb{C}^{2}$. This model falls within a class of GLSMs with unitary gauge group first discussed by Hori and Tong\cite{Hori:2006dk}, that gives an physics explanation of the Pfaffian/Grassmannian correspondence first observed by R{\o}dland\cite{Rodland:2000ab} for a pair of non-birational Calabi-Yau threefolds.

The classical phases are determined by the solutions of the D-term and F-term equations. For our example they read 
\begin{equation}
  \label{dfeq}
  x x^{\dagger}-\sum_k|p^k|^2{\mathrm{id}}_{2\times 2}=\zeta \mathrm{id}_{2\times 2}, \qquad dW=0,
\end{equation}
where $()^{\dagger}$ denotes the conjugate transpose. More generally, the D-terms are given by the moment map equation $\mu(\phi)=\zeta$ with $\mu:V\rightarrow\mathfrak{g}^*$ with $\mathfrak{g}=\mathrm{Lie}(G)$.

The phases of this GLSM are well-understood\cite{Hori:2006dk,Hori:2013gga}. Let $\zeta$ be the FI parameter. The $\zeta\gg0$-phase is a codimension $5$ complete intersection in a Grassmannian
\begin{equation}
  X=\{[x_ix_j]\in G(2,5)|A^{ij}_k[x_ix_j]=0\}, \qquad k=1,\ldots,5.
\end{equation}
The $\zeta\ll0$-phase is a strongly coupled phase with an unbroken $SU(2)$ symmetry given by a determinantal variety
\begin{equation}
  Y=\{p^k\in\mathbb{P}^4|\mathrm{rk}A^{ij}(p)=2\}.
\end{equation}
The two phases are different realisations of the same elliptic curve. So, in this particular model, the two phases are actually isomorphic. In general, two geometric phases of a non-abelian GLSM need not even be birational.

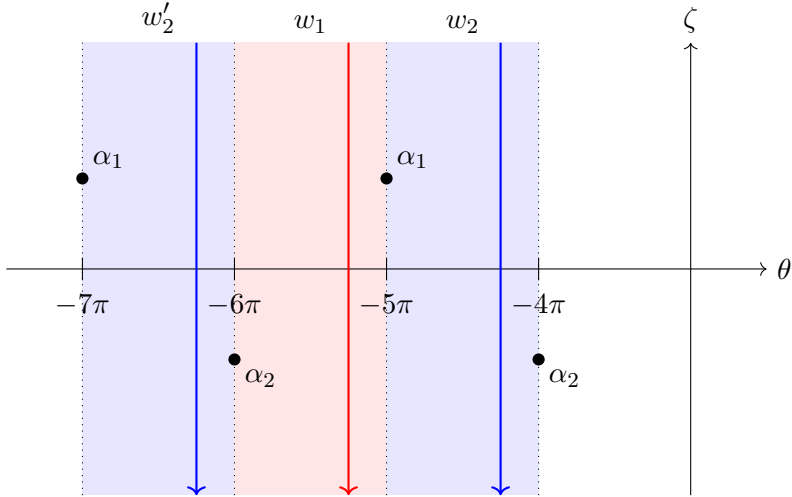
\begin{figure}
  \begin{center}
  \begin{tikzpicture}
  \fill[blue!10!white] (1,-3) rectangle (3,3);
  \fill[red!10!white] (3,-3) rectangle (5,3);
  \fill[blue!10!white] (5,-3) rectangle (7,3);
  \draw (2,3) node [anchor=south] {$w_2'$};
  \draw (4,3) node [anchor=south] {$w_1$};
  \draw (6,3) node [anchor=south] {$w_2$};
  \draw[->] (0,0) -- (10,0) node [anchor=west] {$\theta$};
  \draw[->] (9,-3) -- (9,3) node [anchor=south] {$\zeta$};
  \draw[dotted, thin] (1,-3) -- (1,3);
  \draw[dotted, thin] (3,-3) -- (3,3);
  \draw[dotted, thin] (5,-3) -- (5,3);
  \draw[dotted, thin] (7,-3) -- (7,3);
  \draw (1,-0.15) -- (1,0.15);
  \draw (1,-0.2) node [anchor=north] {$-7\pi$};
   \draw (3,-0.15) -- (3,0.15);
   \draw (3,-0.2) node [anchor=north] {$-6\pi$};
    \draw (5,-0.15) -- (5,0.15);
    \draw (5,-0.2) node [anchor=north] {$-5\pi$};
     \draw (7,-0.15) -- (7,0.15);
     \draw (7,-0.2) node [anchor=north] {$-4\pi$};
     \fill (1,1.2) circle (0.08) node [anchor=south west] {$\alpha_1$};
     \fill (5,1.2) circle (0.08) node [anchor=south west] {$\alpha_1$};
      \fill (3,-1.2) circle (0.08) node [anchor=north west] {$\alpha_2$};
      \fill (7,-1.2) circle (0.08) node [anchor=north west] {$\alpha_2$};
      \draw[->, thick, blue] (2.5,3) -- (2.5,-3);
      \draw[->, thick, red] (4.5,3) -- (4.5,-3);
      \draw[->, thick, blue] (6.5,3) -- (6.5,-3);
  \end{tikzpicture}
  \end{center}
  \caption{Paths and windows for the Pfaffian/Grassmannian elliptic curve GLSM.}\label{fig-paths}
  \end{figure}

Near the phase boundary, there are loci where a Coulomb branch emerges and a maximal $U(1)$ subgroup of $G$ is unbroken. Taking into account quantum corrections, its locus is determined by the critical set of the effective potential
\begin{equation}
  \mathcal{W}_{eff}(\sigma)=-\langle t,\sigma\rangle-\sum_{i=1}^{\mathrm{dim}V}\langle Q_i,\sigma\rangle[\log(\langle Q_i,\sigma\rangle)-1]+\pi i\sum_{\alpha>0}\langle\alpha,\sigma\rangle.
\end{equation}
Here $t=\zeta-i\theta$ are the FI-theta parameters, $\sigma\in\mathfrak{t}_{\mathbb{C}}$ are the scalars of the vector multiplet, $\alpha>0$ are the positive roots of $G$, and $\langle \cdot,\cdot\rangle:\mathfrak{t}^*_{\mathbb{C}}\times\mathfrak{t}_{\mathbb{C}}\rightarrow\mathbb{C}$ with $\mathfrak{t}=\mathrm{Lie}(T)$ where $T\subset G$ is a maximal torus. For our model this becomes
\begin{align}
\mathcal{W}_{eff}(\sigma_1,\sigma_2)  &=-t(\sigma_1+\sigma_2)-5(-\sigma_1-\sigma_2)[\log(-\sigma_1-\sigma_2)-1]\nonumber\\
  &\quad-5\sigma_1[\log(\sigma_1)-1]-5\sigma_2[\log(\sigma_2)-1]+\pi i(\sigma_1-\sigma_2).
\end{align}
The critical set, i.e.~the loci of the Coulomb branch are at
\begin{equation}
  \alpha_1:\: e^{t}=(1+\omega)^5\simeq -e^{2.4}, \qquad \alpha_2:\: e^{t}=(1+\omega^2)^5\simeq e^{-2.4},
\end{equation}
where $\omega=e^{\frac{2\pi i}{5}}$. In other words, the Coulomb branch is located at $\zeta\simeq\pm 2.4$ where the theta angle takes values $0$ or $(\pi$ $\mathrm{mod}\: 2\pi)$. This is depicted in figure~\ref{fig-paths}.

This divides up the (covering of the) moduli space into regions of width $\pi$. One can smoothly go between the phases by choosing a path from the Grassmannian to the Pfaffian phase that avoids the Coulomb branch loci. Paths inside different regions differ by monodromies around the singular points. We note that for this particular example, each path that interpolates between the two phases is homotopic to a straight path with constant $\theta$-angle. This is not true in general\cite{MR3673174,grr}. There are also paths that do not cross the phase boundary. They are homotopic to paths with constant FI-parameter and varying theta angle. This describes large volume monodromies around the limiting points $|\zeta|\rightarrow\infty$ in the moduli space. Since they are easier to deal with, we will not discuss them here and refer to the literature\cite{Herbst:2008jq,Erkinger:2017aaa} for details.  
%%%%%%%%%%%%%%%%%%%%%%%%%%%%%%%%%%%%%%%%%%%%%%%%%%%%%%%%%%%%%%%%%%%%%%%%%%%%%%
\subsection{B-branes and hemisphere partition function}
% branes and MFs, just def and brane factors, modules and wilson line branes, hemisphere , zd2 in phases
A B-brane in a GLSM is characterised by\cite{Herbst:2008jq} the data $\{M,Q,\rho_M,r_* \}$, where $M$ is $\mathrm{Sym}V^*$-module, graded by the the gauge symmetry $\rho_M:G\rightarrow GL(M)$ and the $R$-symmetry $r_*:\mathfrak{u}(1)\rightarrow \mathfrak{gl}(M)$. The free module $M$ has a $\mathbb{Z}_2$-grading, compatible with the gauge and R-symmetry. $Q$ is an odd endomorphism of $M$ that is a gauge-invariant matrix factorisation of the superpotential $W$ of R-charge $1$:
\begin{equation}
  \label{rhordef}
  Q^2(\phi)=W\cdot\mathrm{id}_M, \qquad \rho^{-1}_M(g)Q(g\phi)\rho_M(\phi)=Q(\phi), \qquad \lambda^{r_*}Q(\lambda^R\phi)\lambda^{-r_*}=\lambda Q(\phi),
\end{equation}
where $g\in G$ and $\lambda\in U(1)_R$. We denote the associated category of branes by $MF_G(W)$. To fully specify a GLSM brane we must also specify boundary conditions for the vector multiplet which amounts to finding an admissible\cite{Hori:2013ika} Lagrangian contour $\gamma\subset\mathfrak{t}_{\mathbb{C}}$. The module $M$ can be decomposed in terms of irreducible objects labeled by irreducible representations of $G$ and $R$. Following the GLSM construction, we call these irreducible objects Wilson line branes and denote them by $\mathcal{W}(q^a)_r$ where $q^a$ are the weights of a representation of $G$ and $r$ is the R-charge. Let $S\simeq\mathbb{C}^2$ be the fundamental representation and $\mathrm{det}S$ be the determinantal representation of $U(2)$. They have weights $(q^1,q^2)=\{(1,0),(0,1)\}$ and $(1,1)$, respectively. An example of a constituent of $M$ is $S\otimes\mathrm{det}^2S$.  Assuming it has R-charge $r=3$ this translates as follows into the notation using Wilson line branes:
\begin{equation}
  S\otimes\mathrm{det}^2S\subset M\qquad\leftrightarrow\qquad \mathcal{W}(3,2)_3\oplus\mathcal{W}(2,3)_3
  \end{equation}
Below, we will make use of the notation $R(k)=R\otimes\mathrm{det}^kS$ for some representation $R$ of $U(2)$. 

The hemisphere partition function is defined as\cite{Sugishita:2013jca,Honda:2013uca,Hori:2013ika}
\begin{equation}
  Z_{D^2}(\mathcal{B})=C\int_{\gamma}\mathrm{d}^{\mathrm{rk}G}\sigma \prod_{\alpha>0}\langle\alpha,\sigma\rangle\sinh(\pi\langle\alpha,\sigma\rangle)\prod_{i=1}^{\mathrm{dim}V}\Gamma\left(i\langle Q_i,\sigma\rangle+\frac{R_i}{2}\right)e^{i\langle t,\sigma\rangle}f_{\mathcal{B}}(\sigma),
  \end{equation}
where we refer to the character
\begin{equation}
  f_{\mathcal{B}}(\sigma)=\mathrm{tr}_Me^{i\pi r_*}\rho_M(e^{2\pi\sigma})
\end{equation}
as the brane factor. For our example, the expression reduces to
\begin{align}
  \label{zd2rod5}
  Z_{D^2}(\mathcal{B})&=C\int_{\gamma}\mathrm{d}^2\sigma (\sigma_1-\sigma_2)\sinh(\pi(\sigma_1-\sigma_2))\Gamma\left(-i(\sigma_1+\sigma_2)+1-2\kappa\right)^5\nonumber\\
  &\qquad\quad\cdot\Gamma(i\sigma_1+\kappa)^5\Gamma(i\sigma_2+\kappa)^5e^{it(\sigma_1+\sigma_2)}f_{\mathcal{B}}(\sigma)
\end{align}
To evaluate the partition function in the phases, we can take $\gamma$ to be the real line and we close the integration contour so that the integral converges. In the Grassmannian phase the contributing poles are $\sigma_i=-i(-k_i-\kappa+\epsilon_i)$ with $k_i\in\mathbb{Z}_{\geq 0}$. The integral (\ref{zd2rod5}) can be written as 
\begin{align}
  Z_{D^2}^{\zeta\gg0}(\mathcal{B})=&C'\sum_{k_1,k_2=0}^{\infty}\oint\mathrm{d}^2\epsilon (k_1-k_2-\epsilon_1+\epsilon_2)\frac{(-1)^{k_1+k_2}\pi^{10}\sin\pi(\epsilon_1-\epsilon_2)}{\sin^5\pi\epsilon_1\sin^5\pi\epsilon_2}\nonumber\\
  &\qquad \cdot e^{-t(k_1+k_2-\epsilon_1-\epsilon_2+2\kappa)} \frac{\Gamma(1+k_1+k_2-\epsilon_1-\epsilon_2)^5}{\Gamma(1+k_1-\epsilon_1)^5\Gamma(1+k_2-\epsilon_2)^5}\nonumber\\
  &\qquad \cdot f_{\mathcal{B}}(i(k_1-\epsilon_1+\kappa),i(k_2-\epsilon_2+\kappa)). 
\end{align}
This integral can be rewritten further in terms of an integral over\cite{Hori:2013ika} $X$ including the Chern character of $\mathcal{B}$, the Gamma class and the $I$-function. For our purposes, we will not need this and simply evaluate the residue by computing the $-1$st term of the Laurent expansion in $(\epsilon_1,\epsilon_2)$ of the the integrand. To be consistent with the IR CFT, the R-charge ambiguity has to be chosen to be $\kappa\rightarrow 0$.

In the Pfaffian phase, the contributing poles are
\begin{equation}
  i(\sigma_1+\sigma_2)=1-2\kappa+k-\epsilon_1, \quad i\sigma_2=-\kappa-l+\epsilon_2,\qquad k,l\in\mathbb{Z}_{\geq 0}. 
\end{equation}
The hemisphere partition function evaluated in this phase is
\begin{align}
  Z_{D^2}^{\zeta\ll0}(\mathcal{B})=&C''\sum_{k,l=0}^{\infty}\oint\mathrm{d}^2\epsilon (1+k+2l-\epsilon_1-2\epsilon_2)\frac{(-1)^{l}\pi^{10}\sin\pi(\epsilon_1+2\epsilon_2)}{\sin^5\pi\epsilon_1\sin^5\pi\epsilon_2} \nonumber\\
  &\quad \cdot  e^{t(1+k-\epsilon_1-2\kappa)}\
  \frac{\Gamma(1+k+l-\epsilon_1-\epsilon_2)^5}{\Gamma(1+k-\epsilon_1)^5\Gamma(1+l-\epsilon_2)^5}\nonumber\\
  &\quad\cdot f_{\mathcal{B}}(-i(1+k+l-\epsilon_1-\epsilon_2-\kappa),i(l-\epsilon_2+\kappa)). 
\end{align}
For this phase, we choose $\kappa\rightarrow\frac{1}{2}$. Evaluating the residue yields a divergent sum in $l$. This is typical for strongly coupled phases of non-abelian GLSMs. The sum can be regulated and evaluated explicitly order by order in $k$. In practice, this only works for low orders, making use of Borel summation routines in Mathematica or dedicated software\cite{RISC203,harmonic}.

Upon inserting a valid brane factor (see section~\ref{sec-branes}), the result can be written in terms of periods of the Calabi-Yau that is mirror to both, $X$ and $Y$. The periods are solutions to the Picard-Fuchs equation\cite{Hori:2013gga}
\begin{align}
	\mathcal{L}=\theta^2-z(11\theta^2+11\theta+2)-z^2(\theta+1)^2,
	%\label{eqn:pfrod}
\end{align}
where $\theta=z\frac{d}{dz}$ and we identify $z=-e^{-t}$ as the local coordinate at the large complex structure point, mirror to the $\zeta\gg0$-region in the Grassmannian phase.  
The periods are
\begin{align}
	\varpi_0&=1+3z+19z^2+147z^3+1251z^4+\mathcal{O}(z^5),\\
	(2\pi i)\varpi_1&=\varpi_0\cdot\log(z)+5z+\frac{75}{2}z^2+\frac{1855}{6}z^3+\frac{10875}{4}z^4+\mathcal{O}(z^5),
	%\label{eqn:fiberbasis}
\end{align}
Since we are dealing with an elliptic curve, $X$ and $Y$ are isomorphic, and these are also the periods of $Y$: the Picard-Fuchs operator transforms into itself upon choosing $z\rightarrow -\frac{1}{z}$ and rescaling the holomorphic oneform, which translates to changing the value of the R-charge ambiguity $\kappa$ in the GLSM.

\subsection{Windows and grade restriction}
\label{sec-windows}
% state result, general procedure around monodromy, explain empty brane
By the grade restriction rule\cite{Herbst:2008jq}, depending on the choice of equivalence class of paths, a GLSM brane can only be analytically continued beyond a phase boundary upon restricting the weights of the representation $\rho_M$ to a specific subset associated to the path.  This restriction is often referred to as ``choosing a window''. The grade restriction rule was derived\cite{Herbst:2008jq} by imposing the condition that the boundary effective potential on the Coulomb branch is bounded from below for a given path and a given contour $\gamma$. Once GLSM partition functions have been computed by supersymmetric localisation, it was observed\cite{Hori:2013ika,MR3673174} that the boundary effective potential is encoded in the asymptotics of the hemisphere partition function. Finding the grade restriction rule can then be translated into a convergence condition of the hemisphere partition function.

For our particular model, we expect two infinite families of windows corresponding to the red and blue paths in figure~\ref{fig-paths}. Let $(q^1,q^2)$ be the weights of a $U(2)$-representation. Given a GLSM brane $\mathcal{B}$, these are the gauge charges of the brane encoded in $\rho_M$.  One can show\cite{MR3673174,grr} that the grade-restriction rule for this model is encoded in the following inequalities that have to be satisfied by the weights $q^a$
\begin{eqnarray}
  -2-\frac{5}{4}-\frac{\theta}{2\pi}&<q^1,q^2&<2+\frac{5}{4}-\frac{\theta}{2\pi},\\
  -5-\frac{\theta}{\pi}&<q^1+q^2&<5-\frac{\theta}{\pi},\\
  -\frac{3}{2}&<q^1-q^2&<\frac{3}{2}.
  \end{eqnarray}
%Let $S\cong\mathbb{C}^{2}$ be the fundamental representation of $U(1)$, $\mathrm{det}S$ be the determinantal representation, and $\mathbb{C}$ the trivial representation. Given a representation $R$, we write $R(i)=R\otimes\mathrm{det}^iS$.
The allowed charges can be organised into $U(2)$ representations. For $-6\pi<\theta<-5\pi$ the conditions give
\begin{equation}
  w_1: \quad(q^1,q^2)\in\{\mathbb{C}(1),\mathbb{C}(2),\mathbb{C}(3),\mathbb{C}(4),\mathbb{C}(5),S,S(1),S(2),S(3),S(4)\},
\end{equation}
The window $-5\pi<\theta<-4\pi$ has the allowed charges
\begin{equation}
 w_2:\quad (q^1,q^2)\in\{\mathbb{C},\mathbb{C}(1),\mathbb{C}(2),\mathbb{C}(3),\mathbb{C}(4),S,S(1),S(2),S(3),S(4)\}.
\end{equation}
The windows shifted by $\theta=2\pi n$ are obtained by tensoring the representations above with $\mathrm{det}^{-n}S$. In figure~\ref{fig-paths}, these shifts interpolate between different regions of the same colour. They can be interpreted as large volume monodromies around the limiting points in the respective phases. 

To compute the monodromies around the singular points $\alpha_1,\alpha_2$ we need a third adjacent window which we choose to be $-7\pi<\theta<-6\pi$. This is given by $w_2\otimes\mathrm{det}^1S$:
\begin{equation}
  w_2':\quad (q^1,q^2)\in\{\mathbb{C}(1),\mathbb{C}(2),\mathbb{C}(3),\mathbb{C}(4),\mathbb{C}(5),S(1),S(2),S(3),S(4),S(5)\}.
\end{equation}
In contrast to abelian GLSMs, the asymptotics of the hemisphere partition function also exhibit evidence of grade restriction rules in strongly coupled phases\cite{MR3673174,grr}, consistent with what was observed in the context of equivalences of the categories in the Pfaffian and Grassmannian phases\cite{MR3223878}. For our model this is
\begin{equation}
  -\frac{3}{2}<q^1-q^2<\frac{3}{2}.
\end{equation}
This is independent of the theta angle and the grade restriction rules at the phase boundary are subsets of the grade restriction rule associated to the strongly coupled phase. This gives the following picture of equivalences of categories
\begin{equation}
  \begin{tikzpicture}
  \draw (0,0.5) node {$D(Y)$};
  \draw (4,0.5) node {$D(X)$};
  \draw (2,2) node {$\mathcal{T}_{w_i}$};
  \draw (2,3.5) node {$MF_G(W)$};
  \draw (1.2,2.7) node {$\mathcal{S}$};
  \draw (1.5,3.1) node {$\swgogosubset$};
  \draw (1.5,2.4) node {$\sesubset$};
  \draw[->] (1.7,1.7) -- (0.3,0.8);
  \draw[->] (2.3,1.7) -- (3.7,0.8);
  \draw[->] (2.3,3.3) -- (4,0.8);
  \draw[->] (1.1,2.4) -- (0,0.8);
  \draw (1.3,1.2) node {$\cong$};
  \draw (2.7,1.2) node {$\cong$};
  \end{tikzpicture}
  \end{equation}
Here, $\mathcal{T}_{w_i}$ are subcategories of the category of GLSM branes $MF_G(W)$ where the module $M$ is built out of representations only in the window $w_i$, $\mathcal{S}$ is the subcategory of branes satisfying the grade restriction rule in the strongly coupled phase. 

A generic GLSM brane will not satisfy any grade restriction rule. Branes constructed as GLSM lifts of a brane in a phase can be replaced by objects in the window that describe the same brane in the phase. This is achieved by binding so-called ``empty branes'' using the cone construction. We will discuss this in more detail in section~\ref{sec-empty}.

%%%%%%%%%%%%%%%%%%%%%%%%%%%%%%%%%%%%%%%%%%%%%%%%%%%%%%%%%%%%%%%%%%%%%%
\section{B-branes in the Grassmannian phase}
\label{sec-branes}
In order to obtain the analytic continuation and monodromy matrices, we pick the Grassmannian phase as our reference point and construct a set of GLSM branes that reduce to D0 and D2 branes in the Grassmannian phase. We also identify the empty branes in the Grassmannian phase, which correspond to a set of twisted Lascoux complexes\cite{MR3223878}. 
\subsection{Matrix factorisations for D0 and D2 branes}
% explicit mfs, plus explain problem of W-choice, brane factors and clifford basis, not grade-restricted
We start by constructing the GLSM brane that corresponds to the structure sheaf of $X$. Consider the matrix factorisation
\begin{equation}
  \label{d2q}
  Q_{\mathrm{D}2}=\sum_{k=1}^5 A^{ij}_k[x_ix_j]\eta^k+p^k\overline{\eta}_k,
\end{equation}
where $\eta^k,\overline{\eta}_k$ form a basis of a Clifford algebra with anticommutators $\{\eta^i,\overline{\eta}_j\}=\delta^i_j$ and all others vanishing. If desired, one can construct the explicit matrices and write $Q$ as $32\times 32$-matrix. Consistent with the geometric intuition, the coefficients of $\eta^k$ are the defining equations of the complete intersection in $G(2,5)$. The module $M$ is constructed by choosing a ``vacuum'' $|0\rangle$ and declaring that $\eta_k|0\rangle=0$ so that $M=\bigoplus_{l=0}^5\bigoplus_{i_1<\ldots<i_l}\overline{\eta}_{i_1}\ldots\overline{\eta}_{i_l}|0\rangle$. The conditions (\ref{rhordef}) on gauge invariance and R-charges of $Q$ imply that the $\eta^k$ have gauge charges $(-1,-1)$ and R-charge $1-4\kappa$ and $\overline{\eta}_k$ has gauge charge $(1,1)$ and R-charge $-1+4\kappa$. We write the brane as a complex where the left-most entry is $\prod_{i=1}^5\overline{\eta}_i|0\rangle$ with the maps $A^{ij}_k[x_ix_j]\eta^k$, corresponding to certain block in the matrix $Q$, acting from left to right and the backwards maps $p^k\overline{\eta}_k$ acting from right to left. The right-most term in the complex is always $|0\rangle$. Using our notation for Wilson-line branes, the brane can be characterised by a Koszul-type complex as follows
\begin{equation}
  \label{d2mf}
  \begin{tikzcd}
    \mathcal{W}(q^1,q^2)_{r}\ar[r, shift left]&\ar[l, shift left]\mathcal{W}(q^1+1,q^2+1)^{\oplus 5}_{r+1-4\kappa}%\ar[r, shift left]&\ar[l, shift left]\mathcal{W}(q_1+1,q_2+1)^{\oplus 10}_{r+1-4q}
    %\ar[r, shift left]
    \ldots \ar[r, shift left]&\ar[l, shift left]\mathcal{W}(q^1+5,q^2+5)_{r+5-20\kappa}
    \end{tikzcd}
\end{equation}
Here we have assumed that the leftmost state has gauge charges $(q^1,q^2)$ and R-charge $r$. This fixes the overall normalisation of $\rho_M$ and $r_*$ in (\ref{rhordef}) which is not fixed by the choice of $Q$. In this way, the matrix factorisation (\ref{d2q}) corresponds to infinitely many GLSM branes. Fixing $(q^1,q^2)=0,r=1$, the GLSM brane is equivalent to the Koszul resolution of the structure sheaf $\mathcal{O}_X$ in the Grassmannian phase\footnote{%There is a somewhat unfortunate clash of conventions between the GLSM formulation of branes\cite{Herbst:2008jq} and the mathematics literature.
Most mathematics references use $\mathcal{O}(1)=\mathrm{det}S^{\vee}$ while the physics construction uses $\mathcal{O}(1)=\mathrm{det}S$. This is a matter of perspective: $\mathrm{det}S$ is ample in one phase and anti-ample in the other. The results are compatible up to dualisation which in the physics formulation can be accommodated by making a different choice in building up the Clifford module (eg.~exchange of $\eta$ and $\overline{\eta}$, choosing $|0\rangle$ to be the leftmost state in the complex of Wilson line branes). When talking about GLSM branes, we will adhere to the physics convention, but we will  sometimes use the mathematics convention when quoting results from the literature.}:
\begin{equation}
  \begin{tikzcd}
    0\ar[r]&\mathcal{O}(5)\ar[r]&\mathcal{O}(4)^{\oplus 5}\ldots\ar[r]
    %\mathcal{O}(3)^{\oplus 10}\ar[r]&\mathcal{O}(2)^{\oplus 10}\ar[r]
    &\mathcal{O}(1)^{\oplus 5}\ar[r]&\mathcal{O}\ar[r]&\mathcal{O}_X\ar[r]&0,
    \end{tikzcd}
  \end{equation}
where $\mathcal{O}(k)^{\oplus \binom{5}{j}}=\mathcal{O}(k)\otimes \wedge^jV^{\vee}$.

From (\ref{d2mf}) we deduce the brane factor
\begin{equation}
  f_{\mathrm{D}2}=(-1+e^{2\pi(\sigma_1+\sigma_2)})^5.
\end{equation}
One easily evaluates $Z_{D^2}^{\zeta\gg0}(\mathrm{D}2)=5\varpi_1$. Clearly, (\ref{d2mf}) is not grade-restricted to any window, so it cannot be analytically continued to the Pfaffian phase in a consistent way. Indeed, it describes an empty brane in the Pfaffian phase and one finds $Z_{D^2}^{\zeta\ll0}(\mathrm{D}2)=0$.

Finding a concrete realisation of a simple D0-brane in terms of a GLSM matrix factorisation is more complicated. One approach is to lift a simple geometric brane to the GLSM. The standard option is to intersect the Calabi-Yau with a divisor on the ambient space. This approached has been used, for instance, to construct a D4-brane on the quintic\cite{Herbst:2008jq}. The problem with these branes is that they are very ``far'' out of any grade restriction window in the sense that the range of brane charges is large and therefore they are hard to grade-restrict. The most elementary D0-branes with the smallest charges are some which do not come from divisors on the ambient space. The issue there is that to construct explicit matrix factorisations one has to go to a specific point the the complex structure moduli space where the superpotential has a specific form. In the case of the quintic, this is the Fermat point of the quintic polynomial. For our superpotential (\ref{rodw}) we do not have a Fermat point. However, one can get a suitably nice choice of complex structure, compatible with all genericity conditions of the GLSM, if one replaces the generic $A^{ij}(p)$ by the one defining the mirror of the Pfaffian\cite{boehmthesis}: %{\bf !!! is that really non-degenerate?}
\begin{equation}
  \tilde{A}(p,\psi)=\left(\begin{array}{ccccc}
    0&\psi p^5&p^3&-p^4&-\psi p^1\\
    &0&\psi p^4&p^1&-p^2\\
    &&0&\psi p^2&p^5\\
    &&&0&\psi p^3\\
    &&&&0
  \end{array}\right),
\end{equation}
where $\psi$ corresponds to the complex structure parameter of the mirror and the lower block is determined by the antisymmetry of the matrix. For this example we are allowed to set $\psi=0$ because the rank of this matrix never drops below $2$ if $p\neq 0$. %{\bf !!! check and compare/cite hori-tong}

Then there is a matrix factorisation
\begin{equation}
  Q_{\mathrm{D}0}=\sum_{i=1}^3 x_i^1\eta_i+x_i^2\eta_{i+3}+\ldots,
\end{equation}
where ``$\ldots$'' denotes appropriate factors so that we get a matrix factorisation of $\sum_{i,j}\tilde{A}^{ij}(p,0)[x_ix_j]$. Making an appropriate choice for the gauge and R-charges of  $|0\rangle$, one of the branes associated to this matrix factorisation is characterised by the following complex of Wilson line branes:
\begin{equation}
  \begin{tikzcd}
    \mathcal{W}(0,0)\ar[r,shift left]&\ar[l,shift left]\begin{array}{c}\mathcal{W}(1,0)^{\oplus 3}\\\oplus\\\mathcal{W}(0,1)^{\oplus 3} \end{array}\ar[r, shift left]&\ar[l,shift right]\begin{array}{c}\mathcal{W}(2,0)^{\oplus 3}\\\oplus\\\mathcal{W}(1,1)^{\oplus 9}\\\oplus\\\mathcal{W}(0,2)^{\oplus 3}\end{array}\ar[r,shift left]&\ar[l,shift left]\ldots \ar[r,shift left]&\ar[l,shift left]\mathcal{W}(3,3)
    \end{tikzcd}
\end{equation}
It is easy to see that the organise themselves into irreducible representations of $G=U(2)$ so that the complex can be written as (using $\mathrm{det}S\simeq\mathcal{O}(1)$)
\begin{equation}
  \label{d0mf}
  \begin{tikzcd}
    \mathcal{O}\ar[r,shift left]&\ar[l,shift left] S^{\oplus 3}\ar[r,shift left]&\ar[l,shift left]\begin{array}{c}\mathrm{Sym}^2S^{\oplus 3}\\\oplus\\\mathcal{O}(1)^{\oplus 6}\end{array}\ar[r,shift left]&\ar[l,shift left]\begin{array}{c}\mathrm{Sym}^3S\\\oplus\\ S(2)^{\oplus 8}\end{array}\ar[r,shift left]&\ar[l,shift left]\\
  &  \ar[r,shift left]&\ar[l,shift left]\begin{array}{c}\mathrm{Sym}^2S(1)^{\oplus 3}\\\oplus\\O(2)^{\oplus 6} \end{array}\ar[r,shift left]&\ar[l,shift left]S(2)^{\oplus 3}\ar[r,shift left]&\ar[l,shift left]\mathcal{O}(3)
    \end{tikzcd}
\end{equation}
We have not indicated the R-grading explicitly. Choosing R-charge $r$ for the left-most entry, the R-charge changes by $1-2\kappa$ at each step. Choosing $\kappa=0$ and giving the leftmost entry in the complex even R-charge, the brane factor is
\begin{equation}
  f_{\mathrm{D}0}=(-1+e^{2\pi\sigma_1})^3(-1+e^{2\pi\sigma_2})^3.
\end{equation}
It is easy to show that $Z_{D^2}^{\zeta\gg0}(\mathrm{D}0)=\varpi_0$. Also this brane is not grade restricted to any window. 
%%%%%%%%%%%%%%%%%%%%%%%%%%%%%%%%%%%%%%%%%%%%%%%%%%%%%%%%%%%%%%%%%%%%%%%%
\subsection{Empty branes}
\label{sec-empty}
%cite results and mention lift to lascoux
Empty branes are objects localised on the deleted set associated to the phase. This corresponds to the locus where the GIT quotient determining the phase is not defined. In GLSM language, this are the loci where the D-term equations do not have a solution. For $\zeta>0$ the D-term equation in (\ref{dfeq}) has no solution if the rank of the $2\times 5$-matrix $x=x_i^a$ is less than $2$. We expect the empty branes for this phase to be given in terms of sheaves supported on this locus. A set of suitable exact sequences was proposed in the mathematics literature\cite{MR3223878}. They turn out to be certain twisted Lascoux complexes. Applied to $G(2,5)$, they have the structure
\begin{equation}
  \label{lascoux-schematic}
  X:\quad\begin{tikzcd}
    0\ar[r]&X_5\ar[r]&X_4\ar[r]&X_3\ar[r]&X_2\ar[r]&X_1\ar[r]&0
    \end{tikzcd}
\end{equation}
To grade-restrict branes in the Grassmanian phase, we need four types of complexes:
{\small
\begin{equation}
  \label{lascoux}
  \begin{array}{|c||c|c|c|c|c|}
    \hline
        \textbf{Label}&X_5&X_4&X_3&X_2&X_1\\
        \hline\hline
        A&\mathrm{Sym}^3S^{\vee}(1)\otimes\wedge^5V^{\vee}&\mathrm{Sym}^2S^{\vee}(1)\otimes\wedge^4V^{\vee}&S^{\vee}(1)\otimes\wedge^3V^{\vee}&\mathcal{O}(1)\otimes\wedge^2V^{\vee}&\mathcal{O}\\
        \hline
        B&\mathrm{Sym}^2S^{\vee}(2)\otimes\wedge^5V^{\vee}&S^{\vee}(2)\otimes\wedge^4V^{\vee}&\mathcal{O}(2)\otimes\wedge^3V^{\vee}&\mathcal{O}(1)\otimes V^{\vee}&S^{\vee}\\
        \hline
        C&S^{\vee}(3)\otimes\wedge^5V^{\vee}&\mathcal{O}(3)\otimes\wedge^4V^{\vee}&\mathcal{O}(2)\otimes\wedge^2V^{\vee}&S^{\vee}(1)\otimes V^{\vee}&\mathrm{Sym}^2S^{\vee}\\
        \hline
        D& \mathcal{O}(4)\otimes\wedge^5V^{\vee}&\mathcal{O}(3)\otimes\wedge^3V^{\vee}&S^{\vee}(2)\otimes\wedge^2V^{\vee}&\mathrm{Sym}^2S^{\vee}(1)\otimes V^{\vee}&\mathrm{Sym}^3S^{\vee}\\
        \hline
        \end{array}
\end{equation}}
Each of these complexes can be tensored with $\mathrm{Sym}^kS^{\vee}(l)$ ($k\in\mathbb{Z}_{\geq0},l\in\mathbb{Z}$) to obtain further empty branes.

We need to lift these empty branes to GLSM branes, i.e.~we have to promote them to matrix factorisations of the GLSM superpotential\cite{grr}. 
%The existence of such lift can be argued on general grounds and one can also find an explicit construction in terms of a Heisenberg-Clifford-algebra\cite{grr}, where $\mathrm{Sym}^kS\otimes\wedge^l V$ can be represented as $\oplus_{m_i}\oplus_{n_i}e^{m_1}\ldots e^{m_k}\overline{\eta}_{n_1}\ldots\overline{\eta}_{n_l}|0\rangle$, where the $e^{m_i}$ are the generators of a Heisenberg algebra.
For our purposes, namely computing brane factors and evaluating the hemisphere partition function, we only require the data $(M,\rho_M,r_*)$ but not the maps in the complex encoded in $Q$. All the relevant information is already contained in the data defining the Lascoux complexes. For the sake of making it easy to read off the brane factors, we rewrite the information in (\ref{lascoux}) in terms of Wilson line branes in~\ref{app-lascoux}. There, we also introduce some additional notation to be used below.

%We have not indicated the R-charges in any of the complexes. Each of the $X_i$ in (\ref{lascoux-schematic}) has a definite R-charge that is determined by the condition that $Q$ has R-charge $1$, which induces R-charges of the Clifford matrices $(\eta_i,\overline{\eta}_i)$ and determines the R-charges of the $X_i$. For the Lascoux matrix factoristions there is one subltety: within each complex, there is exactly one map where $\wedge^kV\rightarrow \wedge^{k-2}V$ which would imply a shift in the R-charge by $2$. This however cannot would be inconsistent, as the matrix factorisation would no longer be an odd endomorphism in $M$. To fix this when using the Heisenberg-Clifford construction, we have to correct for this so that the R-charge changes by $\pm 1$ going between $X_i$ and $X_{i+1}$ {\bf !!! check this}. It is easy to show that the brane factors do not depend on the ambiguity $q$ of the R-charge, so we can set $q=0$. Then, the R-charge of each entry of the complexes describing the brane is an integer\footnote{This is not true for general phases of GSLM which can have fractional R-charges.} and the brane factor is only sensitive as to whether this integer is even or odd. Therefore, we often write $\mathcal{W}(q^1,q^2)_{\pm}$ to indicate whether we have an even or odd R-charge.

Given the information from the appendix, the brane factors are:
\begin{align}
  f_A=&1-10 e^{2 \pi  (\sigma_1+\sigma_2)}+ 10 \left(e^{2 \pi  (2 \sigma_1+\sigma_2)}+e^{2 \pi  (\sigma_1+2 \sigma_2)}\right)\nonumber\\
  &-5\left(e^{2 \pi  (3 \sigma_1+\sigma_2)}+e^{2 \pi  (2 \sigma_1+2 \sigma_2)}+e^{2 \pi  (\sigma_1+3 \sigma_2)}\right)\nonumber\\
  &+\left(e^{2 \pi
    (4 \sigma_1+\sigma_2)}+e^{2 \pi  (3 \sigma_1+2 \sigma_2)}+e^{2 \pi  (2 \sigma_1+3 \sigma_2)}+e^{2 \pi  (\sigma_1+4 \sigma_2)}\right)\\
  f_B=&\left(e^{2 \pi  \sigma_1}+e^{2 \pi  \sigma_2}\right)-5 e^{2 \pi  (\sigma_1+\sigma_2)}+10e^{2 \pi  (2 \sigma_1+2 \sigma_2)}\nonumber\\
  &-5 \left(e^{2 \pi  (3 \sigma_1+2 \sigma_2)}+e^{2 \pi  (2 \sigma_1+3 \sigma_2)}\right)\nonumber\\
  &+\left(e^{2 \pi  (4 \sigma_1+2 \sigma_2)}+e^{2 \pi  (3 \sigma_1+3 \sigma_2)}+e^{2 \pi  (2 \sigma_1+4\sigma_2)}\right)\\
  f_C=&\left(e^{4 \pi  \sigma_1}+e^{2 \pi  (\sigma_1+\sigma_2)}+e^{4 \pi  \sigma_2}\right)   -5 \left(e^{2 \pi  (2 \sigma_1+\sigma_2)}+e^{2 \pi  (\sigma_1+2 \sigma_2)}\right)\nonumber\\
  &+10 e^{2 \pi (2 \sigma_1+2 \sigma_2)}-5 e^{2 \pi  (3 \sigma_1+3 \sigma_2)}+\left(e^{2 \pi  (4 \sigma_1+3 \sigma_2)}+e^{2 \pi  (3 \sigma_1+4\sigma_2)}\right)\\
  f_D=&\left(e^{6 \pi  \sigma_1}+e^{2\pi  (2 \sigma_1+\sigma_2)}+e^{2 \pi  (\sigma_1+2 \sigma_2)}+e^{6 \pi  \sigma_2}\right)\nonumber\\
  &-5 \left(e^{2 \pi  (3 \sigma_1+\sigma_2)}+e^{2 \pi  (2 \sigma_1+2 \sigma_2)}+e^{2 \pi  (\sigma_1+3 \sigma_2)}\right)\nonumber\\
  &+10 \left(e^{2 \pi  (3 \sigma_1+2 \sigma_2)}+e^{2 \pi  (2 \sigma_1+3 \sigma_2)}\right)-10 e^{2 \pi  (3 \sigma_1+3 \sigma_2)} +e^{2 \pi  (4 \sigma_1+4\sigma_2)}
  \end{align}
One can check that $Z_{D^2}^{\zeta\gg0}(X)=0$ for $X=\{A,B,C,D\}$. 
%%%%%%%%%%%%%%%%%%%%%%%%%%%%%%%%%%%%%%%%%%%%%%%%%%%%%%%%%%%%%%%%%%%%%%%
\subsection{Grade restriction}
\label{sec-grr}
Grade restriction is a procedure that replaces a GLSM brane that is not in a selected window by an IR equivalent brane that fits the window. This makes use of the triangulated structure of D-brane categories. Using the cone construction, one can bind branes together. In particular, binding any empty branes associated to a phase does not change the brane in the respective phase. We refer to the literature\cite{Herbst:2008jq} for a thorough discussion and focus on the practical aspects. First note that there exist trivial matrix factorisations of the form 
\begin{equation}
  \label{trivmf}
  \begin{tikzcd}
    \mathcal{W}(q_1,q_2)_{r}\ar[r, "1", shift left]&\mathcal{W}(q_1,q_2)_{r+1}\ar[l,
    "W",shift left]
    \end{tikzcd}\qquad \leftrightarrow\qquad Q=\left(\begin{array}{cc}0&1\\W&0 \end{array}\right)=1\eta+W\overline{\eta}
\end{equation}
This describes a configuration with no branes. Whenever such trivial blocks appear in a GLSM brane, they can be removed from the matrix/complex. The other ingredient is the cone construction. We can bind two branes together by turning on a tachyon profile. An open string state $\psi$ between two branes $\mathcal{B}_1$ and $\mathcal{B}_2$ with matrix factorisations $Q_1$ and $Q_2$ is given by a matrix satisfying
\begin{equation}
  Q_1\psi-(-1)^{|\psi|}\psi Q_2=0,\qquad \psi\neq Q_1\phi-(-1)^{|\phi|}\phi Q_2
\end{equation}
for some $\phi$. Here $|\psi|,|\phi|$ is the $\mathbb{Z}_2$-grade. Using this property, it is easy to show that, given a $\mathbb{Z}_2$-odd state $\psi$, one can build a new matrix factorisation $\widetilde{Q}$ as a bound state of the two branes as follows:
\begin{equation}
  \widetilde{Q}=\left(\begin{array}{cc}Q_1&\psi\\0&Q_2 \end{array}\right)
\end{equation}
At the level of complexes $X$ and $Y$ associated to $\mathcal{B}_1$ and $\mathcal{B}_2$, we get the following structure
%\begin{equation}
%  \label{conecomplexes}
%  \begin{tikzcd}
%    &(Y_1)_{r}\ar[r,shift left]&\ar[l,shift left](Y_2)_{r+1}\ar[r,shift left]&\ar[l,shift left](Y_3)_{r+2}\ar[r,shift left]&\ar[l,shift left]\ldots\\
%    (X_1)_{r-1}\ar[r,shift left]\ar[ur, "\psi_1"]&\ar[l,shift left](X_2)_{r}\ar[ur,"\psi_2"]\ar[r,shift left]&\ar[l,shift left](X_3)_{r+1}\ar[ur,"\psi_3"]\ar[r,shift left]&\ar[l,shift left]\ldots&
%    \end{tikzcd}
%\end{equation}
\begin{equation}
  \label{conecomplexes}
  \begin{tikzcd}
    \ldots\ar[r,shift left]&\ar[l,shift left] (X_3)_{r-2}\ar[dr,"\psi_3"]\ar[r,shift left]&\ar[l,shift left](X_2)_{r-1}\ar[dr,"\psi_2"]\ar[r,shift left]&\ar[l,shift left](X_1)_{r}\ar[dr,"\psi_1"]&\\
    &\ldots\ar[r,shift left]&\ar[l,shift left] (Y_3)_{r-1}\ar[r,shift left]&\ar[l,shift left](Y_2)_{r}\ar[r,shift left]&\ar[l,shift left](Y_1)_{r+1}
    \end{tikzcd}
  \end{equation}

Here, $\psi_i$ are the components of $\psi$. Grade restriction now works as follows. Consider an arbitrary GLSM brane that contains a Wilson line brane $\mathcal{W}(q^a)_r$ that is not in any window. Then we can take an empty brane that also contains this Wilson line brane and turn on a tachyon such that there is the identity map between these two Wilson line branes. For example, in the schematic expression (\ref{conecomplexes}), if $(X_1)_r$ is not in the window, we take the complex $Y$ associated to an empty brane such that $X_1=Y_1$ and $\psi_1=\mathrm{id}$. The appearance of the identity assures that the matrix factorisation of the bound state will contain a trivial brane (\ref{trivmf}) that can be removed at the expense of adding some new maps, see \cite[pp.~41-43]{Herbst:2008jq} for a detailed description. This procedure can be iterated until all components that are not in the window have been removed. In practice it is very tedious to compute the maps between the complexes explicitly (see \cite[section~10]{Herbst:2008jq} for some impressive examples). However, since the concrete form of the maps does not enter into the brane factor $f_{\mathcal{B}}$, we do not need the explicit maps. %Assuming they exist\footnote{It would be interesting to know if it is possible to give a general proof, given the properties of empty branes and the requirements of the grade restriction rule.},
The procedure of grade restriction reduces to the combinatorics of placing complexes of Wilson line branes in the correct relative positions. The brane factor of the bound state is then simply the sum of the brane factors of the constituents.  
%%%%%%%%%%%%%%%%%%%%%%%%%%%%%%%%%%%%%%%%%%%%%%%%%%%%%%%%%%%%%%%%%%%%%%%
\section{Analytic continuation, monodromy and modularity}
%general idea to compute analytic continuation and monodromy with brane factors, first analytic continuation, then monodromy at level of brane factors
In this section we compute the analytic continuation matrices $T_w$ for the D0 and D2-branes in the Grassmannian phase, by grade restricting them to a specific window $w$ as indicated in section~\ref{sec-windows}, evaluating the hemisphere partition function in the Pfaffian phase and expressing the result in terms of a basis of central charges of branes in the Pfaffian phase, which in this particular case we can take to be the same as in the Grassmannian phase. Computing the analytic continuation matrices $T_w$ and $T_{w'}$ of two adjacent windows with a singular point $\alpha$ in between the paths in the FI-theta parameter space, we can compute the monodromy matrix $M_{\alpha}$ as $M_{\alpha}=T_w\cdot T_{w'}^{-1}$. 

To get all the information we need, we must grade restrict our branes to the three windows $w_1,w_2$, and $w_2'$. Since the process is rather involved, it is convenient to break up the calculation into smaller building blocks. For this we note the following about the components of the branes (\ref{d2mf}) and (\ref{d0mf}):
\begin{itemize}
\item $\mathcal{W}(5,5)=\mathbb{C}(5)=D_5(1)$ is not in $w_2$.
\item $\mathcal{W}(0,0)=\mathbb{C}=A_1$ is not in $w_1$ and $w_2'$.
\item $\mathcal{W}(1,0)\oplus\mathcal{W}(0,1)=S=B_1$ is not in $w_2'$.
  \item $\mathrm{Sym}^2S=C_1$, $\mathrm{Sym}^2S(1)=C_1(1)$, and $\mathrm{Sym}^3S=D_1$ are not in any window. 
\end{itemize}
Here we have made the connection to the labeling of the entries $X_i$ of the Lascoux branes $X$ in \ref{app-lascoux} that we will also use below. 
To grade restrict, we have to replace the entries that are not in the given window by entries that are. It is not enough to just bind a single suitable empty brane because the empty branes themselves contain entries that are not in any windows. They have to be removed by binding further empty branes until only Wilson line branes fitting the respective window are left. At the level of the brane factors, this amounts to adding brane factors of the respective empty branes with suitable coefficients, where the relative signs are determined by the relative positions of the complexes with respect to the R-charges. It is convenient to break up the calculation into bits that remove the entries that are not in the window. This in particular gives 
\begin{align}
  \label{atoms}
  D_5(1)\rightarrow A_1:\quad& -f_{D(1)}+f_{A}-5f_{B}+5f_{C(1)}\nonumber\\
  A_1\rightarrow D_5(1):\quad&-(-f_{D(1)}+f_{A}-5f_{B}+5f_{C(1)})\nonumber\\
  B_1\rightarrow C_5(2):\quad&-f_B+f_{C(2)}\nonumber\\
  C_5(2)\rightarrow B_1:\quad&-(-f_B+f_{C(2)})\nonumber\\
  C_1:\quad& -f_C\nonumber\\
  C_1(1):\quad& -f_{C(1)}\nonumber\\
  D_1:\quad& -f_D-5f_{C(1)}
\end{align}
The arrows in the first four lines imply that the object on the left, which is not in some of the three windows, gets replaced by the object on the right, which is also not in some of the windows, plus objects that are in all of the windows. These steps only indicate how the respective entries can be removed and how subsequently those parts of the empty branes are removed that are never in any window.  Binding further complexes may be necessary, depending on the concrete window one restricts to.  

For example, the first line in (\ref{atoms}) is to be read as follows. Assume our brane has as its rightmost entry $D_5(1)=\mathcal{W}(5,5)_r$ that is not in the window of our choice. We can remove this entry by taking the complex $D(1)=D\otimes\mathrm{det}S$ in the appendix such that the rightmost entry is $\mathcal{W}(5,5)_{r+1}$ so that the get an identity map $\mathcal{W}(5,5)_r\stackrel{1}{\rightarrow}\mathcal{W}(5,5)_{r+1}$ in the bound state. This entry can be removed. At the level of brane factors, this means that we have to subtract the brane factor $f_{D(1)}$. However, looking at the brane $D(1)$, the entries $D_1(1)=A_5$ and $D_2(2)^{\oplus 5}=B_5^{\oplus 5}$ are never in any window, but we can remove them by binding a copy of empty brane $A$ and five copies of empty brane $B$. However, binding brane $A$ gives us another contribution, $A_4$, that is not in any window. This can be removed by binding the brane $C(1)$. Then we are left with a brane whose components are in every window, plus $\mathcal{W}(0,0)$ that is in a window when $\mathcal{W}(5,5)$ is not. It is also satisfying to see that the inverse operation, namely replacing $\mathcal{W}(0,0)$ with $\mathcal{W}(5,5)$ amounts to a change in overall sign at the level of brane factors. 

Applying the procedure outlined in section~\ref{sec-grr}, the brane factors of the D2 and the D0-brane, grade-restricted to window $w_1$ are
\begin{align}
  f_{\mathrm{D}2}^{w_1}=&f_{\mathrm{D}2}+(-f_{D(1)}+f_{A}-5f_{B}+5f_{C(1)})\nonumber\\
  =&5\big[-(e^{2 \pi  \sigma_1}+e^{2\pi  \sigma_2})+4 e^{2 \pi  (\sigma_1+\sigma_2)}-12 e^{4 \pi  (\sigma_1+\sigma_2)}\nonumber\\
   & +2(e^{2 \pi  (2 \sigma_1+\sigma_2)}+e^{2 \pi  (\sigma_1+2 \sigma_2)})
    +12 e^{6 \pi  (\sigma_1+\sigma_2)}\nonumber\\
   & -2( e^{2 \pi(4  \sigma_1+3  \sigma_2)}+ e^{2 \pi(3  \sigma_1+4  \sigma_2)})-4 e^{8 \pi  (\sigma_1+\sigma_2)}\nonumber\\
    &+( e^{2 \pi  (5 \sigma_1+4 \sigma_2)}+e^{2 \pi  (4\sigma_1+5 \sigma_2)})\big]\\
  f_{\mathrm{D}0}^{w_1}=&f_{\mathrm{D}0}-(-f_{D(1)}+f_{A}-5f_{B}+5f_{C(1)})+3( -f_C)\nonumber\\
  &-(-f_D-5f_{C(1)})+3(-f_{C(1)})\nonumber\\
  =&2(e^{2 \pi  \sigma_1}+e^{2 \pi \sigma_2})-9 e^{2 \pi  (\sigma_1+\sigma_2)}-3(e^{2 \pi  (2 \sigma_1+\sigma_2)}+e^{2 \pi  (\sigma_1+2 \sigma_2)})\nonumber\\
  &+26 e^{4 \pi  (\sigma_1+\sigma_2)}-3(e^{2 \pi(3  \sigma_1+2 \sigma_2)}+ e^{2 \pi(2  \sigma_1+3 \sigma_2)})\nonumber\\
  &-24 e^{6 \pi  (\sigma_1+\sigma_2)}+7( e^{2 \pi(  4\sigma_1+3  \sigma_2)}+ e^{2 \pi( 3 \sigma_1+4  \sigma_2)})+6 e^{8 \pi (\sigma_1+\sigma_2)}\nonumber\\
  &-3(e^{2 \pi (5 \sigma_1+4 \sigma_2)}+ e^{2 \pi  (4 \sigma_1+5 \sigma_2)})+e^{10 \pi  (\sigma_1+\sigma_2)}
  \end{align}
The grade-restricted brane factors for window $w_2$ are
\begin{align}
   f_{\mathrm{D}2}^{w_2}=&f_{\mathrm{D}2}+(-f_{D(1)}+f_{A}-5f_{B}+5f_{C(1)})= f_{\mathrm{D}2}^{w_1}\nonumber\\
 %  =&5\big[(-e^{2 \pi  \sigma_1}-e^{2\pi  \sigma_2})+4e^{2 \pi  (\sigma_1+\sigma_2)}+2(e^{2 \pi  (2 \sigma_1+\sigma_2)}+e^{2 \pi  (\sigma_1+2 \sigma_2)})-12 e^{4 \pi  (\sigma_1+\sigma_2)}\nonumber\\
 %    &+12 e^{6 \pi  (\sigma_1+\sigma_2)} -2(e^{2 \pi (3 \sigma_1+3 \sigma_2)}+e^{2 \pi (3 \sigma_1+4  \sigma_2)})\nonumber\\
 %    &-4 e^{8 \pi  (\sigma_1+\sigma_2)}+( e^{2 \pi  (5 \sigma_1+4 \sigma_2)}+ e^{2 \pi  (4\sigma_1+5 \sigma_2)})\big]\\
    f_{\mathrm{D}0}^{w_2}=&f_{\mathrm{D}0}+3(-f_C)-(-f_D-5f_{C(1)})+3(-f_{C(1)})\nonumber\\
    =&1-3(e^{2 \pi  \sigma_1}+e^{2 \pi  \sigma_2})+6 e^{2 \pi  (\sigma_1+\sigma_2)}+7(e^{2 \pi  (2 \sigma_1+\sigma_2)}+e^{2 \pi  (\sigma_1+2 \sigma_2)})\nonumber\\
    &-24 e^{4 \pi  (\sigma_1+\sigma_2)}-3(e^{2\pi(3  \sigma_1+2   \sigma_2)}+e^{2 \pi (2 \sigma_1+3   \sigma_2)})+26 e^{6 \pi  (\sigma_1+\sigma_2)}\nonumber\\
    &-3(e^{2 \pi(4  \sigma_1+3  \sigma_2)}+ e^{2 \pi (3 \sigma_1+4  \sigma_2)})-9 e^{8 \pi  (\sigma_1+\sigma_2)}\nonumber\\
    &+2(e^{2 \pi  (5 \sigma_1+4 \sigma_2)}+ e^{2 \pi  (4\sigma_1+5 \sigma_2)})
\end{align}
Finally, for the window $w_2'$ we find
\begin{align}
  f_{\mathrm{D}2}^{w_2'}=&f_{\mathrm{D}2}+(-f_{D(1)}+f_{A}-5f_{B}+5f_{C(1)})-5(-f_B+f_{C(2)})\nonumber\\
  =&5\big[-e^{2 \pi  (\sigma_1+\sigma_2)}+2(e^{2 \pi  (2 \sigma_1+\sigma_2)}+e^{2 \pi  (\sigma_1+2 \sigma_2)})-10 e^{4 \pi  (\sigma_1+\sigma_2)}\nonumber\\
    &-5(e^{3 \pi (3 \sigma_1+2  \sigma_2)}+ e^{2 \pi(2  \sigma_1+3  \sigma_2)})+12 e^{6 \pi  (\sigma_1+\sigma_2)}\nonumber\\
    &+3(e^{2 \pi(4  \sigma_1+3  \sigma_2)}+ e^{2\pi( 3 \sigma_1+4  \sigma_2)})-14e^{8 \pi  (\sigma_1+\sigma_2)}\nonumber\\
    &+( e^{2 \pi  (5\sigma_1+4 \sigma_2)}+ e^{2 \pi  (4 \sigma_1+5 \sigma_2)})+5e^{10 \pi  (\sigma_1+\sigma_2)}\nonumber\\
    &-( e^{2 \pi  (6 \sigma_1+5 \sigma_2)}+ e^{2 \pi  (5 \sigma_1+6 \sigma_2)})
    \big]\\
  f_{\mathrm{D}0}^{w_2'}=&f_{\mathrm{D}0}-(-f_{D(1)}+f_{A}-5f_{B}+5f_{C(1)})+5(-f_B+f_{C(2)})-3(-f_B+f_{C(2)})\nonumber\\
  &+3(-f_C)-(-f_D-5f_{C(1)})+3(-f_{C(1)})\nonumber\\
   =& e^{2 \pi  (\sigma_1+\sigma_2)}-3( e^{2 \pi  (2 \sigma_1+\sigma_2)}+ e^{2 \pi  (\sigma_1+2 \sigma_2)})+6 e^{4 \pi  (\sigma_1+\sigma_2)}\nonumber\\
   &+7(e^{2\pi (3 \sigma_1+2  \sigma_2)}+e^{2 \pi(2  \sigma_1+3  \sigma_2)})-24 e^{6 \pi  (\sigma_1+\sigma_2)}\nonumber\\
   &-3( e^{2 \pi (4 \sigma_1+3  \sigma_2)}+ e^{2 \pi (3\sigma_1+4  \sigma_2)})+26 e^{8 \pi  (\sigma_1+\sigma_2)}\nonumber\\
   &-3( e^{2 \pi  (5 \sigma_1+4\sigma_2)}+e^{2 \pi  (4 \sigma_1+5 \sigma_2)})-9e^{10 \pi  (\sigma_1+\sigma_2)}\nonumber\\
   &+2( e^{2 \pi  (6 \sigma_1+5 \sigma_2)}+ e^{2 \pi  (5 \sigma_1+6 \sigma_2)})
  \end{align}
Evaluating the hemisphere partition function in the Grassmannian phase, it is easy to confirm that the result for each window is the same as for the un-grade-restricted branes.

The grade restricted branes are now globally defined along a path in the moduli space specified by the respective window. Hence we can use the brane factors associated to these windows to evaluate the hemisphere partition function in the Pfaffian phase. The result will be the central charge of the analytic continuation of the brane from the Grassmannian phase to the Pfaffian phase. Since the Pfaffian phase has the same periods, we can use the same basis as in the Grassmannian phase. Choosing a basis $\varpi=\varpi^X=\varpi^Y=(\varpi_1,\varpi_0)^T$ in the respective local coordinates, we can compute $\varpi^Y=T_{w_i}\varpi^X$. We find, in agreement with previous results\cite{Knapp:2021vkm}
\begin{equation}
  T_{w_1}=\left(\begin{array}{rr}
  -2&5\\
  5&-13
  \end{array}\right),\quad T_{w_2}=\left(\begin{array}{rr}
  -2&5\\
  -5&12
  \end{array}\right), \quad T_{w_2'}=\left(\begin{array}{rr}
  3&-10\\
  -5&17
  \end{array}\right)
\end{equation}
These matrices have interesting modular properties. All of them are elements of  the congruence subgroup $\Gamma_0(5)$ of the modular group.

Given these results, the monodromies around the singular points $\alpha_1,\alpha_2$ are
\begin{equation}
  M_{\alpha_1}=T_{w_1}\cdot T_{w_2}^{-1}=\left(\begin{array}{rr}
  1&0\\
  -5&1
  \end{array}\right), \qquad
  M_{\alpha_2}=(T_{w_1}\cdot T_{w_2'}^{-1})^{-1}=\left(\begin{array}{rr}
  11&5\\
  -20&-9
  \end{array}\right).
\end{equation}
Both matrices are elements of $\Gamma_1(5)$. In conclusion, analytic continuation is related to $\Gamma_0(5)$, while monodromies are governed by $\Gamma_1(5)$. It would be interesting to understand this better. One possible explanation\footnote{I would like to thank the referee for suggesting this.} for the modular properties of the monodromy matrices is that the FI-theta parameter space of the GLSM is a finite cover of the K\"ahler moduli space of the elliptic curve, implying that the monodromy is a finite-index subgroup of $SL_2(\mathbb{Z})$. 

We note that an alternative way to compute the monodromy matrices is to move the brane around the singular point and back to the Grassmannian phase. This is achieved by taking one of the grade restricted branes and use the empty branes in the Pfaffian phase to grade restrict to the adjacent window and finally to grade-restrict to the window one started with using the empty branes of the Grassmannian phase. In this way one can read off the monodromy transformations directly from the brane factors and there is no need to evaluate the hemisphere partition function. This makes computing monodromy matrices a purely algebraic procedure. Examples of this procedure have been demonstrated to work in abelian GLSMs\cite{Herbst:2008jq,Erkinger:2017aaa}. It is to be expected that this also works in non-abelian GLSMs.
%comment on monodomy with Pfaffian GRR

\section*{Acknowledgments}
I would like to thank the people who have worked with me on GLSMs over the past years: Richard Eager, David Erkinger, Kentaro Hori, Robert Pryor, Mauricio Romo, Emanuel Scheidegger, Thorsten Schimannek, Eric Sharpe, and many others who I had the pleasure to interact with on this and related topics. I thank the Simons Center for Geometry and Physics for hospitality during two workshops in 2023. Thanks to my co-authors\cite{grr} for giving me permission to prepare these notes for the GLSMs@30 proceedings volume. Finally, special thanks go to Eric Sharpe for pitching the idea of a GLSMs@30 workshop and for shouldering most of the organisation. I am supported by the Australian Research Council Discovery Project DP210101502 and the Australian Research Council Future Fellowship FT210100514.

\appendix
\section{Lascoux branes}
\label{app-lascoux}
Here we collect the information of the Lascoux matrix factorisations in terms of complexes of Wilson line branes. The structure of the backwards arrows is actually more complicated than indicated in the diagrams\cite{grr}, but this does not influence the brane factors. The conventions to write the complexes of Wilson line branes are dual to those in the mathematics literature.
\begin{equation}
  A:\quad
  \begin{tikzcd}
    A_1\ar[r, shift left]&A_2^{\oplus 10}\ar[l,shift left]\ar[r,shift left]&A_3^{\oplus 10}\ar[l,shift left]\ar[r,shift left]&A_4^{\oplus 5}\ar[l,shift left]\ar[r,shift left]&A_5\ar[l,shift left],
    \end{tikzcd}
\end{equation}
with
\begin{align}
  A_1&=\mathcal{W}(0,0)\nonumber\\
   A_2&=\mathcal{W}(1,1)\nonumber\\
  A_3&=\mathcal{W}(2,1)\oplus\mathcal{W}(1,2)\nonumber\\
  A_4&=\mathcal{W}(3,1)\oplus\mathcal{W}(2,2)\oplus\mathcal{W}(1,3)\nonumber\\
  A_5&=\mathcal{W}(4,1)\oplus\mathcal{W}(3,2)\oplus\mathcal{W}(2,3)\oplus\mathcal{W}(1,4),
\end{align}
where we assume that all the Wilson line branes in each $A_k$ have the same R-charge, which can either be even or odd. We can tensor the brane by $\mathrm{det}^nS$ which gives the complex $A(n)$, where the brane charges of each component get shifted by $(q_1,q_2)=(n,n)$. All these complexes correspond to empty branes. The other three types of empty branes are as follows. 
\begin{equation}
  B:\quad
  \begin{tikzcd}
    B_1\ar[r, shift left]&B_2^{\oplus 5}\ar[l,shift left]\ar[r,shift left]&B_3^{\oplus 10}\ar[l,shift left]\ar[r,shift left]&B_4^{\oplus 5}\ar[l,shift left]\ar[r,shift left]&B_5\ar[l,shift left],
    \end{tikzcd}
\end{equation}
where
\begin{align}
  B_1&=\mathcal{W}(1,0)\oplus\mathcal{W}(0,1)\nonumber\\
   B_2&=\mathcal{W}(1,1) \nonumber\\
  B_3&=\mathcal{W}(2,2) \nonumber\\
  B_4&=\mathcal{W}(3,2)\oplus\mathcal{W}(2,3) \nonumber\\
   B_5&=\mathcal{W}(4,2)\oplus\mathcal{W}(3,3)\oplus\mathcal{W}(2,4).
  \end{align}
\begin{equation}
  C:\quad
  \begin{tikzcd}
    C_1\ar[r, shift left]&C_2^{\oplus 5}\ar[l,shift left]\ar[r,shift left]&C_3^{\oplus 10}\ar[l,shift left]\ar[r,shift left]&C_4^{\oplus 5}\ar[l,shift left]\ar[r,shift left]&C_5\ar[l,shift left],
    \end{tikzcd}
\end{equation}
with
\begin{align}
C_1&=\mathcal{W}(2,0)\oplus\mathcal{W}(1,1)\oplus\mathcal{W}(0,2)\nonumber\\
C_2&=\mathcal{W}(2,1)\oplus\mathcal{W}(1,2)\nonumber\\
  C_3&=\mathcal{W}(2,2)\nonumber\\
 C_4&=\mathcal{W}(3,3)\nonumber\\
  C_5&=\mathcal{W}(4,3)\oplus\mathcal{W}(3,4).
\end{align}
\begin{equation}
  D:\quad
  \begin{tikzcd}
    D_1\ar[r, shift left]&D_2^{\oplus 5}\ar[l,shift left]\ar[r,shift left]&D_3^{\oplus 10}\ar[l,shift left]\ar[r,shift left]&D_4^{\oplus 10}\ar[l,shift left]\ar[r,shift left]&D_5\ar[l,shift left],
    \end{tikzcd}
\end{equation}
where
\begin{align}
  D_1&=\mathcal{W}(3,0)\oplus\mathcal{W}(2,1)\oplus\mathcal{W}(1,2)\oplus\mathcal{W}(0,3)\nonumber\\
 D_2&=\mathcal{W}(3,1)\oplus\mathcal{W}(2,2)\oplus\mathcal{W}(1,3)\nonumber\\
  D_3&=\mathcal{W}(3,2)\oplus\mathcal{W}(2,3)\nonumber\\
 D_4&=\mathcal{W}(3,3)\nonumber\\
  D_5&=\mathcal{W}(4,4).
\end{align}
We have not indicated the R-charges in any of the complexes explicitly. We set $\kappa=0$ to match with the R-charges of the theory in the IR. The R-charges of the entries $X_i$ of a complex $X$ are always integers and the R-charges of entries $X_i$ and $X_{i+1}$ always differ by $1$. The brane factor  is only sensitive as to whether this integer is even or odd. In our grade restriction calculation we choose the convention that the entry $X_1$ in each complex has even R-charge.

\bibliographystyle{utphys}
\bibliography{proceedings}

\end{document}